\documentclass{desyproc}

\newcommand\etal{et al.}

\begin{document}
%------------------------------------
\title{Tokyo Axion Helioscope}

%for single authors the superscripts are optional
\author{{\slshape M. Minowa$^{1,4}$, Y. Inoue$^{2,4}$, Y. Akimoto$^1$, R. Ohta$^1$, T. Mizumoto$^1$,
A. Yamamoto$^{3,4}$}\\[1ex]
$^1$Department of Physics, School of Science, University of Tokyo\\
$^2$International Center for Elementary Particle Physics, University of Tokyo\\
$^3$High Energy Accelerator Research Organization (KEK),
  1-1~Oho, Tsukuba, Ibaraki 305-0801, Japan\\
$^4$Research Center for the Early Universe (RESCEU), School of Science, University of Tokyo\\
$^{1,2,4}$ 7-3-1 Hongo, Bunkyo-ku, Tokyo 113-0033, Japan}
%\author{{\slshape Axel Lindner$^1$, Konstantin Zioutas$^2$}\\[1ex]
%$^1$DESY, Notketra{\ss}e 85, 22607 Hamburg, Germany\\
%$^2$CERN, 1211 Geneve 23, Switzerland }

% if the proceedings are available online (e.g. at Indico)
% please enter the contribution ID or file_name below for the DOI
%\contribID{32}
\contribID{minowa\_makoto}

% TO THE CONFERENCE EDITORS: 
% please update the following information      
% before sending the template to the authors
% \confID{800}  % if the conference is on Indico uncomment this line
\desyproc{DESY-PROC-2008-02}
\acronym{Patras 2008} % if you want the Acronym in the page footer uncomment this line
\doi  % if there is an online version we will register DOIs

\maketitle

\begin{abstract}
A new search result of the Tokyo axion helioscope is presented. 
The axion helioscope consists of a dedicated cryogen-free 4T superconducting
magnet with an effective length of 2.3\, m and PIN photodiodes as x-ray detectors.
Solar axions, if exist, would be converted into X-ray photons through the inverse
Primakoff process in the magnetic field.
Conversion is coherently enhanced even for
massive axions by filling the conversion region with helium gas.
The present third phase measurement sets a new limit of
$g_{a\gamma\gamma}<\mbox{5.6--13.4}\times10^{-10}\rm GeV^{-1}$
for the axion mass of
$0.84<m_a<1.00\rm\,eV$
at 95\% confidence level.
\end{abstract}

\section{Introduction}
The existence of axion is implied
to solve the so-called strong CP problem
\cite{axion-bible1,axion-bible2,axion-bible3,axion-bible4,axion-bible5}.  
Axions are expected to be produced in solar core through their coupling to photons with energies
of order keV, and the so-called `axion helioscope' technique
may enable us to detect such axions directly
\cite{sikivie1983,bibber1989}.

The differential flux of solar axions
at the Earth is approximated by
\cite{bahcall2004,raffelt2005}
\begin{eqnarray}
  \rm d \Phi_a/\rm d E&=&6.020\times10^{10}[\mathrm{cm^{-2}s^{-1}keV^{-1}}]
  \nonumber\\
  &&{}\times\left(g_{a\gamma\gamma}\over10^{-10}\mathrm{GeV}^{-1}\right)^2
  \left( \frac{E}{1\,\mathrm{keV}}\right)^{2.481}
  \exp \left( -\frac{E}{1.205\,\mathrm{keV}}\right),
  \label{eq:aflux}
\end{eqnarray}
where
$g_{a\gamma\gamma}$ is the axion-photon coupling constant.
Their average energy is 4.2\,keV
reflecting the core temperature of the sun.
Then, they would be coherently converted into X-rays
through the inverse process
in a strong magnetic field at a laboratory.
The conversion rate in a simple case is given by
\begin{equation}
P_{a\to\gamma} = \left( \frac{g_{a\gamma\gamma}B_\bot L}{2}\right)^2
\left[ \frac{\sin(qL/2)}{qL/2}\right]^2,
  \label{eq:prob_plain}
\end{equation}
where
$B_\bot$ is the strength of the transverse magnetic field,
$L$ is the length of the field along the axion path,
$q=(m_\gamma^2-m_a^2)/2E$ is the momentum transfer
by the virtual photon,
$m_a$ is the axion mass,
and 
$m_\gamma$ is the effective mass of the photon
which equals zero in vacuum.

If one can adjust $m_\gamma$ to $m_a$,
coherence will be restored
for non-zero mass axions.
This is achieved by filling the conversion region with gas.
A photon in the X-ray region acquires a positive effective mass
in a medium.
In light gas,
such as hydrogen or helium,
it is well approximated by
\begin{equation}
  m_\gamma=\sqrt{4\pi\alpha N_e\over m_e},
\end{equation}
where $\alpha$ is the fine structure constant,
$m_e$ is the electron mass,
and $N_e$ is the number density of electrons.
We adopted cold helium gas as a dispersion-matching medium.
It is worth noting that
helium remains at gas state even at 5--6\,K,
the operating temperature of our magnet.
Since the bore of the magnet is limited in space,
the easiest way is to keep the gas
at the same temperature as the magnet.
Moreover,
axions as heavy as a few electronvolts
can be reached
with helium gas of only about one atmosphere
at this temperature.

\section{Experimental apparatus}
\begin{wrapfigure}{r}{8cm}
  \includegraphics[scale=0.8]{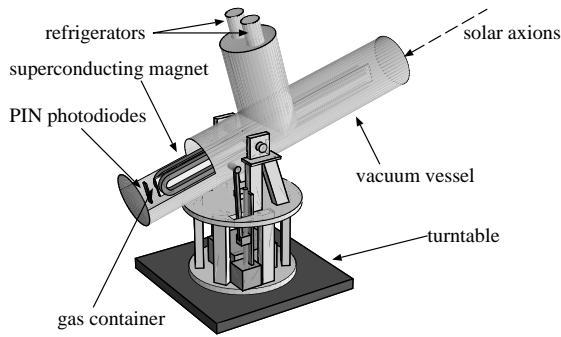}
  \caption{The schematic view of the axion helioscope.}
  \label{fig:sumico}
\end{wrapfigure}

The schematic figure of the axion helioscope is shown
in Fig.~\ref{fig:sumico}.
Its main components are identical to the ones used in the first \cite{sumico1997}
and second phase measurements \cite{sumico2000} of
the Tokyo Axion Helioscope performed in 1997 and 2000, respectively.
It is designed to track the sun in order to achieve
long exposure time.
It consists of
a superconducting magnet, X-ray detectors, a gas container,
and an altazimuth mounting.

The superconducting magnet \cite{sato1997}
consists of two 2.3-m long
race-track shaped coils running parallel
with a 20-mm wide gap between them.
The magnetic field in the gap is 4\,T
perpendicular to the helioscope axis.
The coils are kept at 5--6\,K during operation.
The magnet was made cryogen-free
by making two Gifford-McMahon refrigerators
to cool it directly by conduction,
and is equipped with a persistent current switch.
Thanks to these features, the magnet can be freed from
thick current leads after excitation,
and the magnetic field is very stable for a long period of time
without supplying current.

The container to hold dispersion-matching gas is inserted
in the $20\times92\,\mathrm{mm^2}$
aperture of the magnet.
Its body is made of four 2.3-m long 0.8-mm thick
stainless-steel square pipes
welded side by side to each other.

Sixteen PIN photodiodes, Hamamatsu Photonics S3590-06-SPL,
are used as the X-ray detectors \cite{naniwaPIN},
whose chip sizes are $11\times11\times0.5\rm\,mm^3$ each.
In the present measurement, however, twelve of them are used for the analysis
because four went defective through thermal stresses since the measurement of the previous phase.
The effective area of a photodiode was measured
formerly using a pencil-beam X-ray source,
and found to be larger than $9\times9\,\mathrm{mm^2}$.
It has an inactive surface layer of 
$0.35\,\mu\mathrm{m}$ \cite{akimotoPIN}.

The entire axion detector is constructed in a vacuum vessel
and the vessel is mounted on an altazimuth mount.
Its trackable altitude ranges from $-28^\circ$ to $+28^\circ$
and its azimuthal direction is designed to be limited only
by a limiter which prevents the helioscope from endless rotation.
However, in the present measurement, the azimuthal range is restricted to
about 60$^\circ$ because a cable handling system for its
unmanned operation is not completed yet.

\section{Measurement and Analysis}
From December 2007 through April 2008,
a measurement employing dispersion-matching gas was performed
for 34 photon mass settings with about three days of running time per setting
to scan around 1 eV.

\begin{figure}[hb]
 \centerline{\hbox{%
  \includegraphics[scale=0.6]{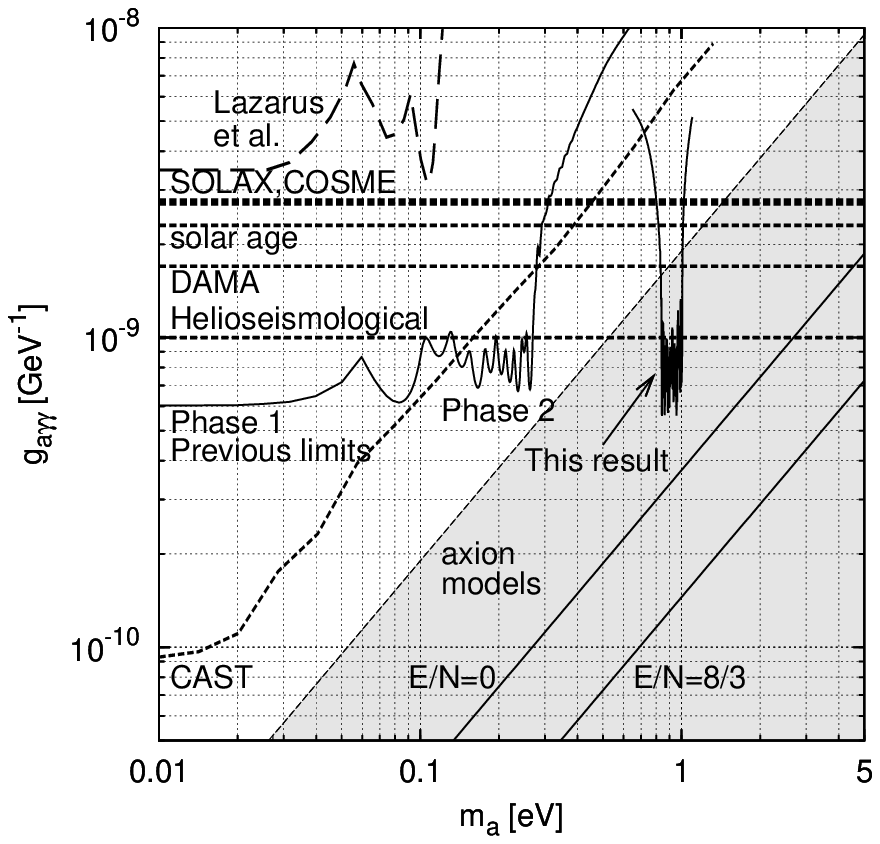}
  \hskip 1cm
  \includegraphics[scale=0.6]{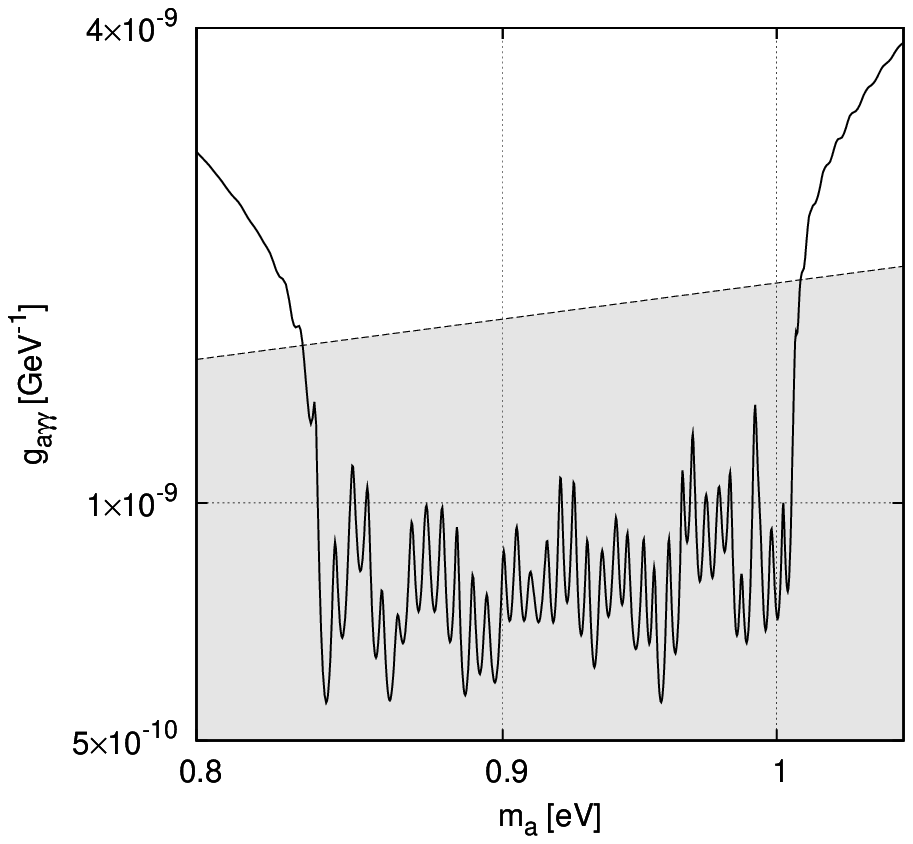}
  }}
\caption{The left figure is the exclusion plot on $g_{a\gamma\gamma}$ to $m_a$.
    The new limit and the previous ones\cite{sumico1997,sumico2000} are plotted in solid lines.
    Dashed lines are explained in the text.
    The hatched area corresponds to the preferred axion models~\cite{GUT_axion}.
    The right figure shows the magnified view of the new limit.
    }
  \label{fig:exclusion}
\end{figure}

Event reduction process is applied 
in the same way as the second phase measurement \cite{sumico2000}.
As a result, no significant excess was seen for any $m_a$,
and thus an upper limit on $g_{a\gamma\gamma}$ at 95\% confidence level
was given.
Fig.~\ref{fig:exclusion} shows
the limit plotted as a function of $m_a$.
Our previous limits from the first \cite{sumico1997}
and the second \cite{sumico2000} phase measurements
and some other bounds are also
plotted in the same figure.
The shown previous limits have been updated using newly measured
inactive surface layer thickness of the PIN photodiode~\cite{akimotoPIN};
the difference is, however, marginal.
The SOLAX~\cite{solax1999}, COSME~\cite{cosme2002} and DAMA~\cite{DAMA2001} are solar axion experiments
which exploit the coherent conversion
on the crystalline planes \cite{Pascos} in a germanium and a NaI detector.
The experiment by Lazarus \etal~\cite{Lazarus} and CAST~\cite{CAST} 
are the same kind of experiments as ours.
The latter utilizes large decommissioned magnets of the LHC at CERN.
Its limit is better than our previous limits by a factor of seven
in low $m_a$ region
due to its larger $B$ and $L$ in Eq.~(\ref{eq:prob_plain}).
In the region $m_a > 0.14\rm\,eV$, however,
our previous and present limits surpass the limit of CAST\footnote{
CAST collaboration showed a preliminary limit in the region $m_a<0.39\, \rm eV$ in the present workshop.}.
The limit $g_{a\gamma\gamma}<2.3\times10^{-9}\rm GeV^{-1}$ is
the solar limit inferred from the solar age consideration
and the limit $g_{a\gamma\gamma}<1\times10^{-9}\rm GeV^{-1}$
is a more stringent limit reported
by Schlattl \etal~\cite{schlattl1999}
based on comparison between
the helioseismological sound-speed profile
and
the standard solar evolution models with energy losses by solar axions.

%
% Conclusion
%
\section{Conclusion}
The axion mass around 1\,eV has been scanned
with an axion helioscope with cold helium gas
as the dispersion-matching medium
in the $4{\rm\,T}\times2.3\rm\,m$ magnetic field,
but no evidence for solar axions was seen.
A new limit on $g_{a\gamma\gamma}$ shown in Fig. \ref{fig:exclusion}
was set for $0.84<m_a<1.00\rm\,eV$.
It is the first result to search for the axion in the
$g_{a\gamma\gamma}$-$m_a$ parameter region of the preferred axion models~\cite{GUT_axion}
with a magnetic helioscope.
Full description of the present result is published in Ref. \cite{sumico2008}.

\section*{Acknowledgments}
The authors thank the former director general of KEK, Professor H. Sugawara,
for his support in the beginning of the helioscope experiment.
This research was partially supported
by the Japanese Ministry of Education, Science, Sports and Culture,
Grant-in-Aid for COE Research and Grant-in-Aid for Scientic Research (B),
and also by the Matsuo Foundation.

% ****************************************************************************
% BIBLIOGRAPHY AREA
% ****************************************************************************

\begin{footnotesize}
% IF YOU DO NOT USE BIBTEX, USE THE FOLLOWING SAMPLE SCHEME FOR THE REFERENCES
% ----------------------------------------------------------------------------

% ----------------------------------------------------------------------------

% IF YOU USE BIBTEX,
% - DELETE THE TEXT BETWEEN THE TWO ABOVE DASHED LINES
% - UNCOMMENT THE NEXT TWO LINES AND REPLACE 'Name_Of_Your_BibFile'

%\bibliographystyle{unsrt}
%\bibliography{Name_Of_Your_BibFile}
% example of Name_Of_Your_BibFile.bib
% @Article{Turcato:2006ch,
%      author    = "Turcato, M.",
%  collaboration = "ZEUS and H1",
%      title     = "Lepton flavour violation and charmonium physics at HERA",
%      journal   = "Nucl. Phys. Proc. Suppl.",
%      volume    = "162",
%      year      = "2006", 
%      pages     = "283-287",
%      SLACcitation  = "%%CITATION = NUPHZ,162,283;%%"
% }
% 
% @Unpublished{Gogitidze:2007du,
%      author    = "Gogitidze, N.",
%  collaboration = "H1", 
%      title     = "Prompt photons and particle momentum distributions at
%                   HERA", 
%      year      = "2007",
%      note    = "hep-ex/0701033",
%      SLACcitation  = "%%CITATION = HEP-EX 0701033;%%"
% }

\end{footnotesize}

% ****************************************************************************
% END OF BIBLIOGRAPHY AREA
% ****************************************************************************

\end{document}